\documentclass[epjST]{svjour}
\usepackage{graphicx}
\usepackage{epstopdf}
\usepackage{subfigure}
\usepackage{amsmath}
\usepackage{amssymb}
\begin{document}
\title{Quantum Monte Carlo study of the indirect Pauli exclusion effect in Bose-Fermi mixtures}
\author{G. Bertaina\inst{1}\fnmsep\thanks{\email{gianluca.bertaina@unimi.it}} \and A. Guidini\inst{2} \and P. Pieri\inst{2} }
\institute{Dipartimento di Fisica, Universit\`a degli Studi di Milano, via Celoria 16, I-20133 Milano, Italy \and School of Science and Technology, Physics Division, 
University of Camerino, Via Madonna delle Carceri 9, I-62032 Camerino, Italy}
\abstract{
We study the momentum distributions of a three-dimensional resonant Bose-Fermi mixture in the molecular limit at zero temperature. For concentration of the bosons with respect to the fermions less or equal to one, each boson is bound to a fermion and the system is composed of fermionic molecules plus excess fermions. Not only the bosonic condensate fraction goes to zero, signaling a quantum phase transition towards a normal phase, but a finite region of low momenta is depleted, depending on the concentration. This phenomenon is named indirect Pauli exclusion effect and is demonstrated via Fixed-Node Diffusion Monte Carlo simulations and T-matrix calculations.
} 
\maketitle
\section{Introduction}
\label{intro}
The use of Feshbach resonances to control effective interactions in alkali atoms has allowed for reaching regimes where pairing effects are strongest. In two-component Fermi mixtures, tuning interactions beyond the unitarity limit has experimentally opened the vast field of the BCS-BEC crossover, of molecular condensates and quantum phase transitions driven by polarization \cite{exppol1,exppol2}. The combined effect of strong interactions and intermediate polarizations has renewed interest in non-conventional pairing mechanisms, such as in the Sarma-Liu-Wilczek phase \cite{sarma,liu}. In its simplest form this phase represents a mixture of dimers and excess atoms where the excess fermions occupy the low momenta states, while fermions inside the molecules are pushed to occupy high momentum states and  the center-of-mass momentum of the bosonic molecules is zero. This results in a depleted momentum distribution of the minority fermions, up to the Fermi momentum of the excess fermions, with a sharp step to the pairing region. An analogy can be driven for mixtures of bosons and single-component fermions in the strongly interacting regime, with larger fermionic than bosonic densities $n_F>n_B$ (see \cite{Fra10,Fra12,Fra13,Guidini14,Ber13} and references therein). There one expects the emergence of a Fermi-Fermi mixture of excess and molecular fermions \cite{Ber13}, where excess fermions occupy low momenta states, while fermions and bosons in the molecules occupy higher momenta. When $n_B<n_F-n_B$, the bosonic momentum distribution clearly manifests this depletion effect (See Fig. \ref{fig:TMAscale}), which can be called indirect Pauli exclusion induced by pairing. Differently than in two-component Fermi mixtures, in the strongly interacting regime of Bose-Fermi mixtures the dimers must have finite momentum, since they are fermions, so that a sharp jump in the bosonic momentum distribution is not present. 

In \cite{Guidini14} we showed how this interesting phenomenon can be dealt with T-matrix formalism, discussing the validity of suitable approximate expressions for the momentum distribution functions. We also compared those results with a dedicated Fixed-Node Diffusion Monte Carlo (FN-DMC) simulation for a small concentration of bosons $x=n_B/n_F=0.175$. Here we present a more extensive comparison at different concentrations.  Throughout this paper we only consider $g=(k_F a_{BF})^{-1}=3$, where $a_{BF}$ is the Bose-Fermi scattering length and $n_F=k_F^3/6\pi^2$, as a representative example of the strongly interacting regime, and we restrict to the equal masses case $m_B=m_F$.

\section{Momentum distributions from T-matrix calculations}
\label{sec:TMA}
We adopt the formalism described in \cite{Guidini14} for the calculation of the momentum distributions using a T-matrix approximation (TMA) for the self-energies. We also show the results coming from an almost analytical strong-coupling asymptotic  simplification of the TMA expressions (eqs. 26 and 35 of \cite{Guidini14}, labeled ``asymptotic'' in the figures).

\begin{figure} [tb]
\centering
\includegraphics[width=0.6\textwidth]{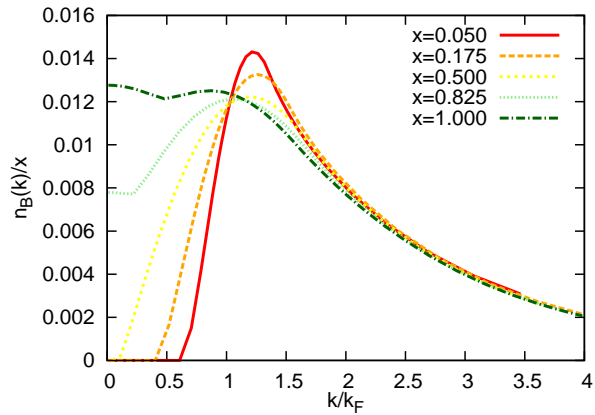}
\caption{T-matrix evaluation of the bosonic momentum distributions divided by the corresponding concentration $x$.}
\label{fig:TMAscale}
\end{figure}

A particularly clear presentation of the bosonic momentum distributions is shown in Fig. \ref{fig:TMAscale}, where we divide each curve by the corresponding concentration $x$, so that the integral is the same. One sees that the high momentum tail is practically universal, being asymptotically equal to the square of the Fourier transform of the two-body molecular wave function. For momenta $k\lesssim 2 k_F$, the distributions strongly depend on $x$. Please note that, even if for $x>0.5$ one has a finite $n_B(k=0)$, there is no BEC, since $n_B(k=0)$ does not scale with the volume of the system.

%

\section{Quantum Monte Carlo methodology}
\label{sec:QMC}
We now present a comparison between the T-matrix results and FN-DMC simulations. The FN-DMC method performs an imaginary-time evolution of the initial trial wave function $\Psi_T$, with a suitable mapping to a drift-diffusion stochastic process complemented by a birth-and-death branching process for an efficient management of the statistical weight. To circumvent the fermionic sign problem, the nodal surface of the solution is constrained to be the same as for $\Psi_T$. Simulations are carried out in a cubic box of volume $L^3=N_{\rm F}/n_{\rm F}$ with periodic boundary conditions.  Details of the implementation are the same as in \cite{Guidini14,Ber13}.

Bose symmetry of the trial wave function is crucial when calculating momentum distributions, since these are calculated as mixed estimators $n_{\rm B}(k)=\langle \Psi_T|\hat{n}_k|\Psi_0\rangle/\langle \Psi_T|\Psi_0\rangle$, where $\hat{n}_k$ is the number operator in momentum space averaged over momentum direction, and $\Psi_0$ is the long-(imaginary)-time evolution of $\Psi_T$. An explicit symmetrization over all permutations of the bosons of the molecular wave function $\Phi_A^{\rm MS}$ \cite{Ber13} is not feasible since the number of terms to be summed scales as the factorial of the number of bosons $N_{\rm B}$. As in \cite{Guidini14}, we instead implicitly perform the symmetrization over the bosonic coordinates within the molecular orbitals, with an approach similar to the one employed in \cite{cazorla2009} in the context of solid $^4$He.

We write the guiding wave function as $\Psi_T({\bf R})=~\Phi_S({\bf R})\Phi_A({\bf R})$. $\Phi_S$ is a usual symmetric Jastrow function $\Phi_S({\bf R})=\prod_{i^\prime j^\prime}f_{\rm BB}(r_{i^\prime j^\prime})\prod_{i j}f_{\rm FF}(r_{ij})$, where the unprimed (primed) coordinates refer to fermions (bosons) and two-body spherically symmetric correlation functions of the interparticle distance are introduced. $f_{\rm BB}$ is the solution of the two-body Bose-Bose problem with $f_{\rm BB}^\prime(L/2)=0$; $f_{\rm FF}$ is described below. Antisymmetrization of the fermionic coordinates is provided by the use of a generalized Slater determinant of the following form:
\begin{equation}\label{eq:psiA-MS-approx}
 \tilde{\Phi}_A^{\rm MS}({\bf R})=\left|\begin{matrix}
                        \varphi_{K_1}(1,{\bf R}_{\rm B}) & \cdots & \varphi_{K_1}(N_{\rm F},{\bf R}_{\rm B}) \\
			\vdots                   &\ddots & \vdots                  \\
			\varphi_{K_{N_{\rm M}}}(1,{\bf R}_{\rm B}) &\cdots & \varphi_{K_{N_{\rm M}}}(N_{\rm F},{\bf R}_{\rm B}) \\
			\psi_{k_1}(1)          &\cdots & \psi_{k_1}(N_{\rm F})     \\
			\vdots                           &\ddots & \vdots                  \\
			\psi_{k_{N_R}}(1)        &\cdots & \psi_{k_{N_R}}(N_{\rm F})
                       \end{matrix}\right|\;,
\end{equation}
where the molecular orbitals are a sum $\varphi_{K_\alpha}(i,{\bf R}_{\rm B})=\sum_{i^{\prime}}\varphi_{K_\alpha}(i,i^\prime)$ of the two-body orbitals  $\varphi_{K_\alpha}(i,i^\prime)=f_{\rm B}(|{\bf r}_i-{\bf r}_{i^\prime}|)\exp{(i {\bf K}_\alpha ({\bf r}_i+{\bf r}_{i^\prime})/2)}$, which consist of the relative-motion orbitals $f_{\rm B}$ times the molecular center-of-mass plane waves with $|K_\alpha|\le P_{\rm CF}$, and $n_{\rm CF}=n_B=P_{\rm CF}^3/6\pi^2$, while for the unpaired fermions $|k_\alpha|\le k_{\rm UF}$, with $n_{\rm UF}=n_F-n_B=k_{\rm UF}^3/6\pi^2$. The functions $f_{\rm B}$ are chosen to be the bound solution of the two-body Bose-Fermi problem, asymptotically modified to respect boundary conditions.

By expanding the determinant it is easy to get convinced that \eqref{eq:psiA-MS-approx} contains all possible permutations of the bosons among the $N_B$ fermionic molecules, but it contains also additional terms where many different molecular orbitals $\varphi$ are occupied by the same boson and the remaining bosons' coordinates do not explicitly appear, thus effectively belonging to plane waves at zero momentum. These spurious terms thus tend to increase the bosonic condensate. They moreover correspond to  the clustering of many fermions close to a single boson at a distance of order of $a_{\rm BF}$; they are then significant near resonance, where the molecular orbitals are very loose, while hopefully they are strongly suppressed in the molecular limit due to the Pauli principle, which forbids the formation of fermion clusters.
 We have tried to suppress further these spurious terms, by introducing a very short-range repulsive Jastrow factor between fermions as described in \cite{Guidini14}, obtaining a reduction of the variance of the observables.
 
FN-DMC simulations of  \eqref{eq:psiA-MS-approx} are quite heavy, given the $O[N_F^3]$ scaling due to the required inversion of the matrix in \eqref{eq:psiA-MS-approx}, times the $O[N_B]$ scaling due to the calculation of the symmetrized orbitals. We have used $N_B=7$ bosons and $N_F=7\div 40$ fermions, set so as to consider closed shells and reduce finite-size effects.

\section{Comparison of Quantum Monte Carlo and T-matrix results}

\begin{figure}[tb]
\centering
\subfigure[$x=0.029$]{
\includegraphics[width=0.48\textwidth]{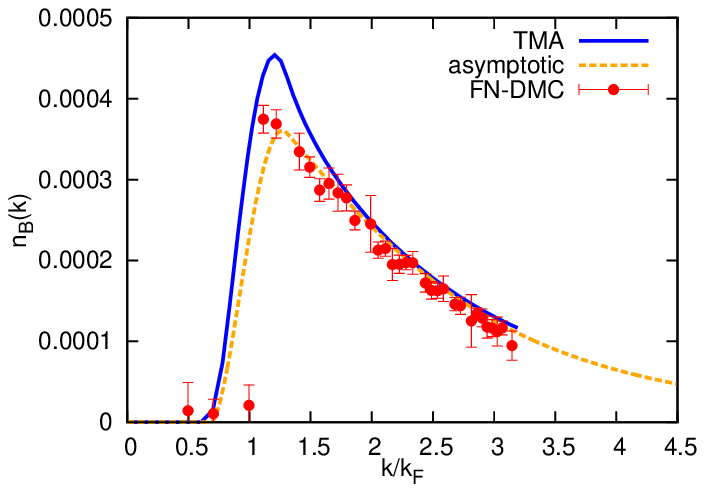}}
\subfigure[$x=0.175$]{
\includegraphics[width=0.48\textwidth]{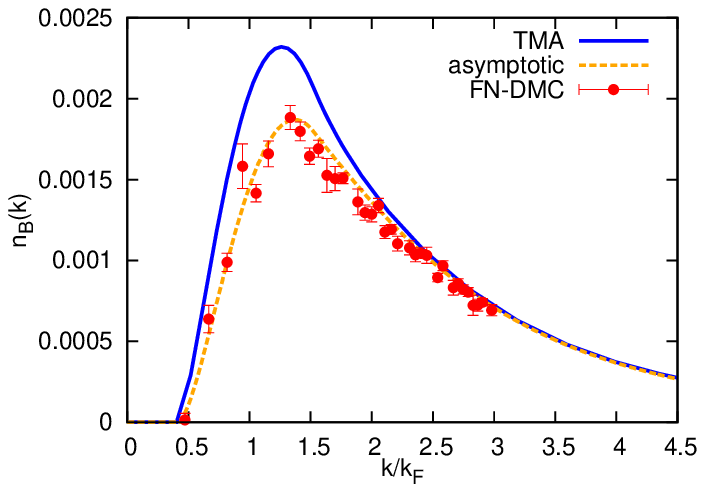}\label{fig:0175}}
\subfigure[$x=0.500$]{
\includegraphics[width=0.48\textwidth]{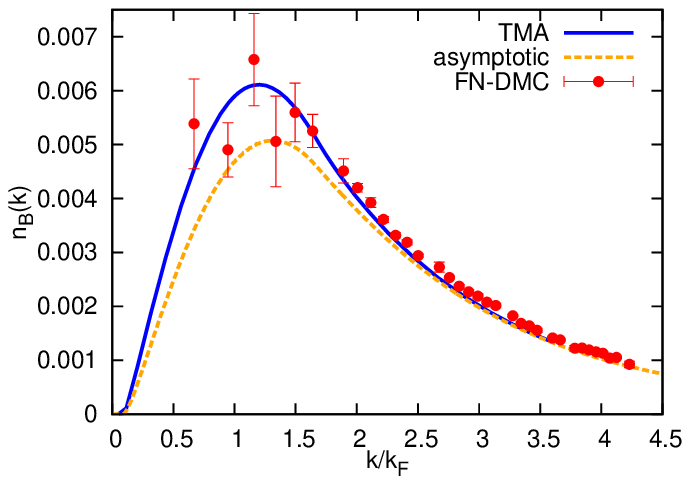}}
\subfigure[$x=1.000$]{
\includegraphics[width=0.48\textwidth]{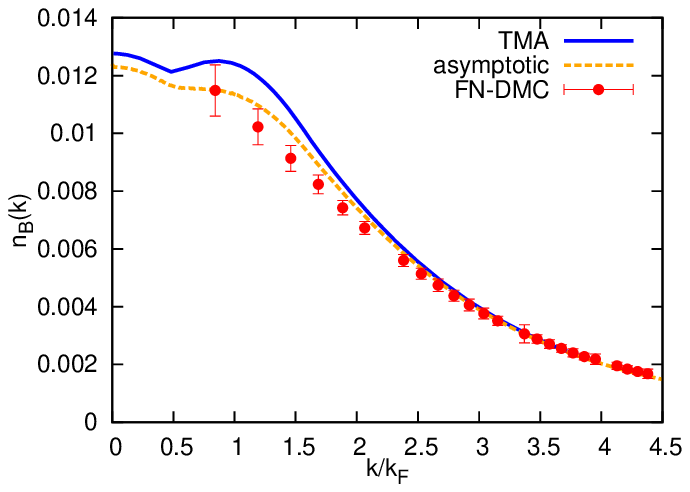}}
\caption{Comparison of T-matrix and FN-DMC results for the bosonic momentum distributions at various concentrations and $g=3$. The FN-DMC data in (a) correspond to the impurity problem with $N_F=34$. This explains the relatively sharp jump at $k=k_F$. Data in (b) are taken from \cite{Guidini14}.}
\label{fig:allx}       
\end{figure}

\begin{figure} [tb]
\centering
\includegraphics[width=0.5\textwidth]{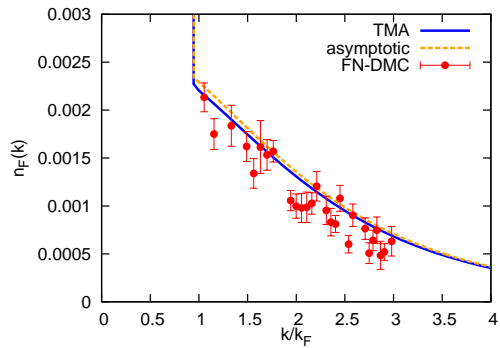}
\caption{Comparison between T-matrix results, asymptotic expression (eq. 35 of \cite{Guidini14}) and FN-DMC calculations for the fermionic momentum distribution at $g=3$ and $x=0.175$. For $k<k_U\simeq0.94$ the distribution goes to $\sim1$ with both the techniques.}
\label{fig:0175f}
\end{figure}

In Fig. \ref{fig:allx} we compare the FN-DMC and T-matrix results for $n_{\rm B}(k)$ at $g=3$, for the concentrations $x=0.029, 0.175, 0.500, 1.000$. We only show the mixed estimate of $n_B(k)$, even if it is biased by $\Psi_T$; although a common way of reducing this bias is by extrapolation using the Variational Monte Carlo estimator, this is prevented here since the difference between $\Psi_0$ and $\Psi_T$ is not small when calculating the condensate. Actually we observed that a significant initial transient time in the FN-DMC simulations was spent in extinguishing the spurious contributions giving fermionic clusters, which are energetically costly and increase the bosonic condensate. Finite-size effects are stronger in panels (c) and (d) where the numbers of fermions are $14$ and $7$. This does not affect the high momenta tail, which is mainly determined by two-body properties. On the other side, finite-size effects generally tend to increase occupancy of low momenta. Given also the discussion on the symmetrization of the trial wave function, we can thus cautiously say that the true $n_B(k)$ is probably smaller in this region. For all concentrations FN-DMC was able to deplete the condensate fraction down to a value smaller than $n_0=0.002$. The FN-DMC calculations confirm the suppression of $n_B(k)$  at low $k$ for $x<0.5$. For $x=0.5$ the small considered system size does not allow to probe a large number of momenta at low $k$ and the large variance prevents definitive assertions for $k<k_F$, while at $x=1$ the results are partially quantitatively agreeing with the T-matrix calculation. The FN-DMC calculations are in good agreement with the  T-matrix results at high momenta ($k \gtrsim 2$). Some deviations occur in the intermediate $k-$region, where the FN-DMC results seem closer to the strong coupling curves rather than the full T-matrix results. This may indicate that the relative motion molecular orbital $f_{\rm B}$, close resembling the bare molecule wave function, strongly affects the nodal surface, and refining its parametrization would probably increase the occupancy of intermediate momenta. The very structure of wave function \eqref{eq:psiA-MS-approx}, being essentially of strong-coupling type, may affect this momentum region.

In Fig. \ref{fig:0175f} we also show the fermionic momentum distribution for $k>k_F$ at $x=0.175$. The agreement between FN-DMC and T-matrix results is reasonable. By reducing the statistical error, we expect relevant finite-size effects to become apparent, which are now smaller than the error bars.  

\section{Concluding remarks}
We have demonstrated the low occupancy of bosonic momenta at low concentration with two different methods. Although results are encouraging and should hopefully stimulate experimental investigation, overcoming the biased nature of the FN-DMC estimators would be a major advancement. Future work will employ zero temperature path integral methods, which generically remove the initial wave function bias.

\begin{acknowledgement}
We acknowledge the CINECA Award LI03s-AccelQMC (2014) for the availability of high performance computing resources and support.
\end{acknowledgement}

\end{document}